-----------------------------------------------------------------------
\documentstyle[preprint,aps]{revtex}

\begin{document}

\title{Low-energy direct muon transfer from H
to Ne$^{10+}$, S$^{16+}$ and Ar$^{18+}$ using two-state
close-coupling approximation
to Faddeev-Hahn-type equation}


\author{Renat A. Sultanov$^{1,2}$\thanks{e-mail:
renat@rainbow.physics.tamu.edu} and Sadhan K.
Adhikari$^{1}$\thanks{e-mail: adhikari@ift.unesp.br}}

\address{$^1$Instituto de F\'\i sica Te\'orica,
Universidade Estadual Paulista,\\
01405-900 S\~{a}o Paulo, S\~{a}o Paulo, Brazil\\
$^2$Department of Physics, Texas A \& M University,\footnote{Present
address}
\\College Station, Texas
77843}

\date{\today}
\maketitle

\begin{abstract}

We perform three-body calculation of direct muon-transfer rates from
thermalized muonic hydrogen isotopes to bare nuclei Ne$^{10+}$,
S$^{16+}$, and Ar$^{18+}$ employing integro-differential
Faddeev-Hahn-type equations in configuration space with
two-state close-coupling
approximation scheme. All Coulomb potentials including the strong
final-state Coulomb repulsion are treated exactly.  A long-range
polarization potential is included in the elastic channel to take into
account the high polarizability of the muonic hydrogen. The transfer
rates so calculated are in good agreement with recent experiments. We
find that the muon is captured predominantly in the $n=6$, 9, and 10
states of muonic Ne$^{10+}$, S$^{16+}$, and Ar$^{18+}$, respectively.


\end{abstract}


\section{Introduction}

A theoretical description of three-body charge-transfer reactions in
physics is both challenging and interesting and is a subject of
active research \cite{1,2,3,5,6,7,8,9}, as the study of this  fundamental
problem  
provides a
testing ground of the underlying physical model. This is even more so when
the mass of the exchanged charged particle is not negligible compared to
the other two and when there is a strong final-state Coulomb interaction.
The first feature invalidates the commonly-used simplifying
Born-Oppenheimer-type
approximation scheme \cite{2} and calls for a detailed three-body
description. The
second feature demands a proper dynamical treatment of the final-state
Coulomb interaction. 

The direct muon ($\mu$) transfer rates at low energies from the
muonic-hydrogen
isotopes
$^1$H$_\mu$ ($p\mu$) and $^2$H$_\mu$ ($d\mu$) to bare 
nuclei 
\mbox{X}$^{Z+}$ of large charge $Z$:
\begin{equation}\label{eqq}
(\mbox{H}_\mu)_{1s} + \mbox{X}^{Z+}\ \rightarrow \
(\mbox{X}_\mu)^{(Z-1)+} + \mbox{H}^+
\end{equation}
have been the subject of many experimental investigations
\cite{pr97,jetp66,nci69,nci67,np80,nci67a,pra95,jpb93,pr98}, where H
stands
for the hydrogen isotope proton ($p$) or deuteron ($d$) and X$^{Z+}$
stands for the target nuclei. Investigations of the low-energy
muon-transfer reactions in collisions of muonic hydrogen H$_\mu$ with
nuclei X$^{Z+}$ are of importance for the muon-catalyzed fusion cycle
\cite{r}. 
 Although, there have been many
experiments of muon transfer with nuclei of large charge $Z$,
the theoretical investigations are limited
to nuclei X$^{Z+}$ with $Z =1$ \cite{8} and $Z=2$ \cite{9}. 
The theoretical description of these reactions at
low energies has both the
complicating features mentioned above and becomes extremely complicated as
the charge $Z$ of the nuclei increases. This is due to the strong
electromagnetic field of the nuclei simultaneously responsible for a large
initial-state polarization and a strong final-state Coulomb interaction.
Also, the large mass of the transferred muon compared to the electron
leads to
additional difficulties  compared to the electron-transfer reaction where
a
Born-Oppenheimer-type approximation is efficient \cite{2}.  In addition, a
large
number of open channels in the muon-transfer reactions even at zero energy
complicates the theoretical
treatment.  It is difficult to incorporate these effects properly in a
dynamical three-body calculation. This is why there are {\it no} published
work to date on a three-body dynamical 
calculation of these muon-transfer rates for nuclei with charge $Z > 3$.
These aspects demand a careful three-body quantum mechanical treatment
with exact inclusion of the final-state Coulomb interaction.

The recent theoretical activities in charge transfer are centered  around
problems with much weaker polarization and Coulomb interactions compared
to those encountered in muon transfer involving bare nuclei like argon or
sulphur. In electron-hydrogen-atom \cite{1}, deuteron-hydrogen-atom
\cite{2},  
positron-hydrogen-atom \cite{5}, electron-positronium-atom
\cite{6},  hydrogen-positronium-atom \cite{7} and H-$\mu$-H \cite{8}
systems 
the final-state Coulomb interactions are zero compared to 
$17e^2/\rho$ in the case of $p$-$\mu$-Ar$^{18+}$ considered here, where
$e$
is the electronic charge and 
$\rho$ is the radial separation in the final state. In proton-deuteron
\cite{3} and H-$\mu$-He \cite{9} systems there is a small
final-state Coulomb repulsion, e.g., $e^2/\rho$.
 The
initial-state polarization is also much smaller in those studies
\cite{1,2,3,5,6,7,8,9}. The large charge of the bare nuclei  complicates
substantially the analysis
of  the present muon transfer problems.

Recently, we presented a theoretical formalism for the study of such
muon-transfer reactions using two-component integro-differential
Faddeev-Hahn-type equations \cite{hahn68} in configuration space with
close-coupling
approximation scheme \cite{pra2000},
and applied it to the study of muon transfer from
muonic hydrogen isotopes to He$^{2+}$ and Li$^{3+}$ \cite{jphysb}, and
C$^{6+}$ and O$^{8+}$ \cite{pra2001}. 
The transfer rates $\lambda_Z$
calculated there
are
in good agreement with experiments, whenever available.   
The formalism of Ref. \cite{pra2001} seems to be appropriate  for the
study of the problem of
muon transfer from muonic hydrogen isotopes to nuclei with
large charge.

Among such  transfer reactions, muon transfers from muonic hydrogen
isotopes   to Ne$^{10+}$, S$^{16+}$, and Ar$^{18+}$
have been the center of active interest  to different
experimental groups. 
There have been several  experiments
for muon transfer from $p\mu$ to argon which
yielded transfer rates centered around the following five different
values: $(16.3 \pm 0.9)\times 10^{10}$s$^{-1}$
 \cite{pr97},
        $(12.0 \pm 1.9)\times 10^{10}$s$^{-1}$
 \cite{jetp66},
        $(14.6 \pm 1.4)\times 10^{10}$s$^{-1}$
 \cite{nci69},
        $(35 \pm 6)\times 10^{10}$s$^{-1}$
 \cite{nci67}
and     $(98 \pm 15)\times 10^{10}$s$^{-1}$
 \cite{np80}.
For  muon transfer from $d\mu$ to
argon the experimental transfer rates are
$(8.6  \pm 0.4)\times 10^{10}$s$^{-1}$
 \cite{pr97}
and  9.4 $\times 10^{10}$s$^{-1}$
 \cite{nci67a}. The experimental situation in the case of argon is quite
controversial with widely different values for the transfer rates,
specially in the case of $p\mu$. Despite this intense experimental
activity in these 
muon-transfer reactions there are no quantum dynamical calculations valid
at
low
energies for these transfer rates. 
In view of the above interest and  controversy, in
this paper
we undertake the
challenging three-body study of direct muon transfer from $p\mu$ and
$d\mu$ to argon using the formulation of Ref. \cite{pra2001}.
To test our approach, in addition, we apply it to study muon transfer from
hydrogen isotopes to neon and sulphur where experimental results are
available.  
In the case of Sulphur  
the experimental transfer rate from $p\mu$ is  $ 8.9\times
10^{10}$s$^{-1}$, whereas that from $d\mu$ is $11 \times 10^{10}$s$^{-1}$
\cite{pra95}.
In the case of neon
the experimental transfer rate 
from $d\mu$ is $10.1 \times 10^{10}$s$^{-1}$
\cite{jpb93}.

Although there is a very large number of open channels in these problems,
for a given nuclei the muon is transferred predominantly to a few (muonic)  
atomic labels of the heavy nuclei \mbox{X}$^{Z+}$ \cite{zpa,ann77,ger63}.  
For example, it was first noted by Gershtein \cite{ger63} and reconfirmed
later \cite{pra2001} that the muon is captured mostly in the $n=4$ states
of C$^{6+}$, and $n=5$ states of O$^{8+}$ \cite{pr98,pra2001}.  Also these
transfers take place mostly to the final muonic-atomic states with low
angular momenta and transfer rates are negligible for muonic atomic states
with angular momenta $l>2$. 
Semiclassical description of these muon
transfer reactions has been very useful in explaining many qualitative and
quantitative features \cite{ann77,cla}. Using a semiclassical model based 
on potential curves of the two-centered $p\mu$X$^{Z+}$ system, 
Holzwarth et al. \cite{zpa} demonstrated that muon transfer to F$^{9+}$
takes place essentially to the $n=6$ level of the (F$_\mu$)$^{8+}$ 
system. They
also showed that transfer to the nearby levels of the   (F$_\mu$)$^{8+}$  
atom is
negligible compared to the $n=6$ level. 
 From similar  consideration of
semiclassical barrier penetration, Haff et al. \cite{ann77} derived the
following formula for the state $n$ of the (X$_\mu$)$^{(Z-1)+}$ atom to
which  
the muon from H$_\mu$ is predominantly transferred in reaction (\ref{eqq})
\begin{equation}\label{for} n={\cal I}n
\left[\frac{Z(1+2Z^{1/2})}{1+2/Z^{1/2}} \right]^{1/2}, \end{equation}
where ${\cal I}n$ denotes the integral part. For C$^{6+}$, O$^{8+}$ and 
F$^{9+}$
this formula leads to $n=4$, 5, and 6, respectively, in agreement with
calculation
\cite{zpa,ger63}.   The formula (\ref{for}) is expected to
work even better for heavier nuclei Ne$^{10+}$, S$^{16+}$ and Ar$^{18+}$
where the channelization to a final specific state $n$ should be more
perfect with muon transfer to nearby $n$ states heavily suppressed.

The correct dynamical formulation should include all open
transfer channels and we included them in a previous study on
muon transfer with light nuclear targets \cite{jphysb}.  However, it is
quite impossible now to treat even this reduced number of open transfer
channels in a quantum  calculation with heavier targets due to
convergence difficulties in the presence of the large final-state Coulomb
interaction mentioned above. Hence, in the present treatment we use a
two-channel model to calculate transfer to a single final state, where we
include the elastic and one transfer channel.  Different sets of equations
are used for the different final states.  
Eventually, the total transfer
rate is calculated by summing the different contributions. 
 After
calculation we find that in both cases ($p\mu$ and $d\mu$) the muon is
captured predominantly in the $n=6$ state of Ne$^{10+}$, $n=9$ state of
S$^{16+}$, and $n=10$ state of Ar$^{18+}$ in complete agreement with
formula (\ref{for}) and in this pioneering theoretical study we present
results for these
cases. 
The transfer is highly suppressed to other values of $n$ and
higher $l$ states.

In
the Faddeev-Hahn-type equation \cite{hahn68}, the wave function is broken
up into
components with proper asymptotic behavior in different physical channels.
Consequently, these wave-function components are much smoother functions
of configuration-space variables than the Schr\"odinger wave function
which is the sum of all these components. Hence, this approach
simplifies the solution procedure and  the correct asymptotic
behavior of the solution in different channels can be incorporated
easily. In addition, these 
equations 
allow us to introduce explicitly a polarization potential in the initial
channel.  All Coulomb potentials including the strong 
final-state Coulomb repulsion  are treated exactly (without approximation
or
parameters) in this formalism. The effect of strong polarization of the
muonic hydrogen by the bare nuclei is accounted for by a polarization
potential with a cut off parameter. By a proper inclusion of the
polarization potential in the intermediate region, the present transfer
rates are found to be essentially independent of this parameter.  The
correct inclusion of the final-state Coulomb interaction has the advantage
of building in the proper asymptotic behavior of the wave function in a
low-order close-coupling type approximation \cite{fb99a,fb99b}. Hence as
in
Ref. \cite{pra2001} we make a two-state close-coupling approximation to
the
Faddeev-Hahn-type equation in the present study and find that a numerical
solution using the present scheme leads to very encouraging agreement with
recent experimental transfer rates.

The theoretical framework for the present study is based on the formalism
developed in Refs. \cite{pra2000,jphysb,pra2001} which was used for the
study of
muon transfer from muonic hydrogen atoms to lighter charged nuclei H$^+$,
He$^{2+}$, Li$^{3+}$, C$^{6+}$, and O$^{8+}$.  
 In the dynamical equations in
Refs. \cite{jphysb,pra2001} the final-state Coulomb interaction in the
transfer
channel is treated exactly 
without approximation. In addition, as in Ref. \cite{pra2001}, here we
explicitly include a polarization potential in the elastic channel. 
The presence of the strong
Coulomb interaction and the associated large polarization make the present
calculational scheme far more complicated  numerically
compared to those of Refs. \cite{jphysb,pra2001}. 
 In a
coupled-channel approach for atomic processes, the coupling to infinite
number of p-wave states is responsible for generating the polarization
potential \cite{mott65,prs60}.  As it is impossible to include all such
states in a numerical scheme, the commonly accepted procedure is to
replace these coupling terms by the polarization potential as in
Ref. \cite{rmp62}. 

In Sec. II we present a brief account of the theoretical formulation. In
Sec. III we report the present numerical results and finally, in Sec. IV
we present some concluding remarks.

\section{Theoretical Formulation}
A detailed account of the theory has already appeared
\cite{pra2000,jphysb,pra2001}. For the sake of completeness we report
here a summary of the theoretical development.
 Here we describe the dynamical equations we use
based on the close-coupling
approximation to Faddeev-Hahn-type two-component equations \cite{jphysb}. 
We use units $e=\hbar=m_\mu=1$, where $m_\mu$($e$) is
the muonic  mass  (charge), and denote the heavy nucleus
X$^{Z+}$ by ${\sf 1}$, the hydrogen isotope(s)
by ${\sf 2}$ and muon by ${\sf 3}$.
Below the three-body breakup threshold, following 
two-cluster asymptotic configurations
are possible in the system {\sf 123}:  $({\sf 23})\ -\ {\sf 1}$ and
$({\sf 13})\ -\ {\sf 2}$. These two configurations 
  correspond to two distinct physical channels,  denoted by  
1 and 2, respectively. 
These configurations  
are  determined by the Jacobi coordinates
$(\vec r_{j3}, \vec \rho_k)$:
$\vec r_{13} = \vec r_3 - \vec r_1,
\hspace{6mm} \vec \rho_2 =
(\vec r_3 + m_1\vec r_1) / (1 + m_1) - \vec r_2$,
$\vec r_{23} = \vec r_3 - \vec r_2,
\hspace{6mm} \vec \rho_1 =
(\vec r_3 + m_2\vec r_2) / (1 + m_2) - \vec r_1$, where
$\vec r_{i}$, $m_{i}$ ($i=1, 2, 3,$) are coordinates and
masses of the particle $i$, respectively.

Let us introduce
the total three-body wave function as a sum of two components
\begin{equation}
\Psi(\vec r_1, \vec r_2, \vec r_3) \ =\  \Psi_1 (\vec r_{23},\vec \rho_1)
\ + \ \Psi_2 (\vec r_{13},\vec \rho_2)
\label{eq:total2}
\end{equation}
where $\Psi_1 (\vec r_{23},\vec \rho_1)$
is quadratically integrable over the variable
$\vec r_{23}$, and  $\Psi_2 (\vec r_{13},\vec \rho_2)$ over
$\vec r_{13}$. The components $\Psi_1$ and $\Psi_2$ carry the
asymptotic
boundary condition for channels 1 and 2, respectively.
The second component is responsible for
pure Coulomb interaction in the final state.
These components satisfy the following 
set of two coupled equations
\begin{eqnarray}\label{eq:1a}
[E - (H_0 + V_{23}(\vec r_{23})) - \ U_{\scriptsize \mbox{pol}} 
(\vec \rho_1)]\Psi_1 (\vec
r_{23}, \vec \rho_1) &=&
[(V_{23}(\vec r_{23})  + V_{12}(\vec r_{12})) - 
U_{\scriptsize \mbox C}\ (\vec \rho_2)]
\Psi_2 (\vec r_{13}, \vec \rho_2)\; \\
\vspace{6mm}
[E - (H_0 + V_{13}(\vec r_{13})) - \ U_{\scriptsize \mbox C}\ (\vec
\rho_2)]\Psi_2 (\vec r_{13}, \vec \rho_2) &=&
[(V_{13} (\vec r_{13})+ V_{12}(\vec r_{12}))
-U_{\scriptsize \mbox{pol}}(\vec
\rho_1)]\Psi_1 (\vec r_{23}, \vec \rho_1)
\label{eq:1}
\end{eqnarray}
where $E$ is the center-of-mass energy, $H_0$ the total kinetic energy
operator,  $V_{ij} (\vec r_{ij})$
the pair potential $(i \not= j = 1, 2, 3)$, 
$U_{\scriptsize \mbox C}$ the final-state Coulomb  interaction given
by
\begin{equation}
U_{\scriptsize \mbox C}(\vec \rho_2) = \frac{(Z - 1)Z'}{\rho_2}
\end{equation}
with $Z$ the charge of the heavy nuclei  and $Z'(=1)$  the
charge of the hydrogen  isotope. 
Here $U_{\scriptsize \mbox{pol}}$ is the
polarization potential given by 
 \cite{ger63}
\begin{equation}
U_{\scriptsize \mbox{pol}}(\vec \rho_1) = -\frac{9Z^2}{4\rho_1^4}
\hskip 0.5cm
\mbox{for}
\hskip 0.5cm
\rho_1 > \Lambda
\label{eq:pol}
\end{equation} 
and zero otherwise. The value of the cut-off parameter $\Lambda$ has to be
chosen appropriately.
By adding equations (\ref{eq:1a}) and (\ref{eq:1}) we find that they are
equivalent to the Schr\"odinger equation.

Distortion potentials are useful in phenomenological models of scattering
in atomic physics \cite{rmp62}. Although, unnecessary in a complete
solution of
the Schr\"odinger equation, they facilitate the numerical effort in a
simplified model.  We have included the proper polarization potential in
the initial channel. Although, a polarization potential exists in the
final rearrangement channel, the most important interaction in this
channel is the Coulomb repulsion which has been exactly included in our
description.

Because of the strong final-state Coulomb repulsion, it is very difficult
to solve the multichannel model equations based on Eqs. (\ref{eq:1a}) and
(\ref{eq:1}). Hence,   for solving (\ref{eq:1a}) and  (\ref{eq:1}) we
expand the wave function
components in terms of bound states in initial and final channels,
and project these equations on these bound states. The expansion of the
wave function is given by
\begin{equation}
\Psi_1(\vec r_{23}, \vec \rho_{1}) \approx
\frac{f_{1s}^{(1)}(\rho_1)}{\rho_1}
R_{1s,\mu_1}^{(Z')}(|\vec r_{23}|)/4\pi
\label{eq:expan1}
\end{equation}
\begin{equation}
\Psi_2(\vec r_{13}, \vec \rho_{2}) \approx
\frac{f_{nl{\cal L}}^{(2)}(\rho_2)}{\rho_2}
R_{nl,\mu_2}^{(Z)}(|\vec r_{13}|)
\left \{ Y_{\cal L}(\hat \rho_2) \otimes
Y_l(\hat r_{13}) \right \}_{00}
\label{eq:expan2}
\end{equation}
where $nl{\cal L}$ are quantum numbers of the
three-body final-state,
$\mu_1=m_3m_2/(m_3+m_2)$, $\mu_2=m_3m_1/(m_3+m_1)$,
$Y_{lm}$'s
the  spherical harmonics,
$R_{nl,\mu_i}^{(Z)}(|\vec r|)$ the radial part of the
hydrogen-like bound-state wave function for reduced mass $\mu_i$
and charge $Z$,
$f_{1s}^{(1)}(\rho_1)$ and
$f_{nl{\cal L}}^{(2)}(\rho_2)$ the unknown expansion coefficients.
This prescription is similar to that adopted in the close-coupling
approximation. After a proper angular momentum projection,
the  set of two-coupled integro-differential equations
for the unknown expansion functions can be written as 
\begin{eqnarray}
\left[ (k_1^{(1)})^2\ +\ \frac{\partial^2}
{\partial \rho_1^2}\ - \
2M_1U_{\scriptsize \mbox{pol}} (\vec \rho_1) \right]
f_{1s}^{(1)}(\rho_1) =
g_1\sqrt{(2{\cal L} + 1)}
\int_{0}^{\infty} d \rho_{2}
f_{nl{\cal L}}^{(2)}(\rho_{2})
\; \nonumber \\
\times \int_{0}^{\pi}d \omega \sin\omega
R_{1s,\mu_1}^{(Z')}(|\vec{r}_{23}|)
\left(-\frac{Z'}{|\vec {r}_{23}|} + \frac{Z}{|\vec {r}_{12}|}
 - U_{\scriptsize \mbox C} (\vec \rho_2) \right)
R_{nl,\mu_2}^{(Z)}(|\vec{r}_{13}|)
\;\nonumber \\
\times \rho_1 \rho_2
C_{{\cal L} 0l0}^{00}
Y_{lm}(\nu_{2}, \pi)/\sqrt{4\pi}
\label{eq:most1}
\end{eqnarray}
\begin{eqnarray}
\left[ (k^{(2)}_n)^2\ +\ \frac{\partial^2}
{\partial \rho_2^2}\ -\
\frac{{\cal L} ({\cal L} + 1)}{\rho_2^2} -
2M_2U_{\scriptsize \mbox C} (\vec \rho_2)
\right] f_{nl{\cal L}}^{(2)}(\rho_2) = g_2
\sqrt{(2{\cal L }+ 1)}
\; \nonumber \\
\times \int_{0}^{\infty} d \rho_{1}
f_{1s}^{(1)}(\rho_{1})\int_{0}^{\pi}
d \omega \sin\omega
R_{nl,\mu_2}^{(Z)}(|\vec{r}_{13}|)
\left(-\frac{Z}{|\vec {r}_{13}|} + \frac{Z}{|\vec {r}_{12}|}
- U_{\scriptsize \mbox{pol}} (\vec \rho_1)\right)
\;\nonumber \\
\times R_{1s,\mu_1}^{(Z')}(|\vec{r}_{23}|)
\rho_{2} \rho_1
C_{{\cal L }0l0}^{00}
Y_{lm}(\nu_{1}, \pi)/\sqrt{4\pi}\;.
\label{eq:most2}
\end{eqnarray}
Here 
$k_1^{(1)}   = \sqrt{2M_1{(E- E_{1s}^{(2)})}}$,
$k_n^{(2)} = \sqrt{2M_2{(E- E_n^{(1)})}}$
with
$M_1^{-1}= m_1^{-1} + (1 + m_2)^{-1}$ and
$M_2^{-1}= m_2^{-1} + (1 + m_1)^{-1}$,
$E_n^{(j)}$ is the binding energy of pair $(j3)$ and
$g_j=4\pi M_j/\gamma^{3}$ ($j = 1, 2$), 
$\gamma=1-m_1m_2/((1+m_1)(1+m_2))$,
$C_{{\cal L} 0lm}^{Lm}$ the Clebsch-Gordon coefficient,
$L$ the total angular momentum,
$\omega$ the angle between the Jacobi coordinates
$\vec \rho_1$ and $\vec \rho_2$, $\nu_1$ the angle between 
$\vec r_{23}$ and $\vec \rho_1$ and $\nu_{2}$ the angle
between $\vec r_{13}$ and $\vec \rho_{2}$.

To find unique solution to
(\ref{eq:most1}) and (\ref{eq:most2}),
appropriate boundary conditions are to be considered.  We impose the
usual condition of regularity at the origin 
$f_{1s}^{(1)}(0)\mathop{\mbox{\large$=$}}0$ and
$f_{nl\cal L}^{(2)}(0)\mathop{\mbox{\large$=$}}0$.
Also
for the present scattering  problem with $1 +(23)$ as the initial state,
in the asymptotic region, two solutions to
(\ref{eq:most1}) and (\ref{eq:most2}) satisfy the following boundary
conditions
\begin{eqnarray}
f_{1s}^{(1)}(\rho_1)
&\mathop{\mbox{\large$\sim$}}\limits_{\rho_1 \rightarrow + \infty}&
\sin(k^{(1)}_1\rho_1) + {\it K}_{11}^{nl}\cos(k^{(1)}_1\rho_1)
\\
f_{nl\cal L}^{(2)}(\rho_2)
&\mathop{\mbox{\large$\sim$}}\limits_{\rho_2 \rightarrow + \infty}&
\sqrt{v_1 / v_2}{\it K}_{12}^{nl}
\cos(k^{(2)}_1\rho_2 - \eta / 2k^{(2)}_1 \ln2k^{(2)}_1\rho_2
-\pi{\cal L}/2)
\label{eq:cond88}
\end{eqnarray}
where 
$\it K_{ij}^{nl}$ are appropriate coefficients.
For scattering
with ${\sf 2} + ({\sf 13})$ as the initial state, we have the following
conditions
\begin{eqnarray}
f_{1s}^{(1)}(\rho_1)
\mathop{\mbox{\large$\sim$}}\limits_{\rho_1 \rightarrow + \infty}
\sqrt{v_2 / v_1}{\it K}_{21}^{nl}\cos(k^{(1)}_1\rho_1)\\
f_{nl\cal L}^{(2)}(\rho_2)
\mathop{\mbox{\large$\sim$}}\limits_{\rho_2 \rightarrow + \infty}
\sin(k^{(2)}_1\rho_2 - \eta / 2k^{(2)}_1 \ln2k^{(2)}_1\rho_2
-\pi{\cal L}/2)\nonumber \\
+{\it K}_{22}^{nl} \cos(k^{(2)}_1\rho_2 -
\eta / 2k^{(2)}_1 \ln2k^{(2)}_1\rho_2 - \pi{\cal L}/2)
\label{eq:cond8888}
\end{eqnarray}
where $v_i$ ($i=1,2$) is the velocity in channel $i$. 
The Coulomb parameter in the second transfer channel
is $\eta = 2M_2(Z-1)/k^{(2)}_n$  \cite{mott65}.
The coefficients $\it K_{ij}^{nl}$ are obtained from the
numerical solution of the Faddeev-Hahn-type equations.
The cross sections are given by
\begin{eqnarray}
\sigma^{\scriptsize \mbox{tr}}_{1s\rightarrow nl}\ =\
\frac{4\pi(2L+1)}{k^{(1)2}}\frac{({\it K}_{12}^{nl})^2}
{(D - 1)^2 + ({\it K}_{11}^{nl} + {\it K}_{22}^{nl})^2}
\end{eqnarray}
where $D = {\it K}_{11}^{nl}{\it K}_{22}^{nl}
-{\it K}_{12}^{nl}{\it K}_{21}^{nl}$. When $k^{(1)} \rightarrow 0$,
$\sigma^{\scriptsize \mbox{tr}}_{1s\rightarrow nl} \sim 1/k_1^{(1)}$.
The transfer rates  are defined by 
\begin{equation}
\lambda^{\scriptsize \mbox{tr}}_{1s\rightarrow nl}\ 
=\ \sigma^{\scriptsize \mbox{tr}}_{1s\rightarrow nl} v_1 N_0
\label{eq:tr1}
\end{equation}
where $v_1$ is the relative velocity of the incident
fragments and 
$N_0$ the liquid-hydrogen density chosen here
as $4.25\times10^{22}$ $\mbox{cm}^{-3}$. We note that
$\lambda^{\scriptsize \mbox{tr}}(k^{(1)}\rightarrow 0) \sim $ const.
In our model the total muon transfer rate is
\begin{equation}
\lambda_{\scriptsize \mbox{tot}}^{\scriptsize \mbox{tr}} = 
\sum_{nl} \lambda^{\scriptsize \mbox{tr}}_{1s\rightarrow nl}.
\label{eq:tr2}
\end{equation}

\section{Numerical Results}

We employ muonic atomic unit: distances are measured in units of $a_\mu$,
where $a_\mu$ is the radius of muonic hydrogen atom.  The
integro-differential equations are solved by discretizing them into a
linear system of equations as in Refs. \cite{jphysb,pra2001}. 
As we are concerned with the low-energy limit, only the total
angular momentum $L=0$ is taken into account.  Even at zero incident
energy, the transfer channels are open and their wave functions are
rapidly oscillating Coulomb waves. A large number of discretization points
is needed for achieving convergence.

First Eqs.  (\ref{eq:most1}) and (\ref{eq:most2}) are solved setting  the
polarization potential
to zero. 
The numerical convergence for the system of equations including the
elastic and a transfer channel (at a time) is obtained after moderate
effort. Finally, the total
transfer cross section is calculated by adding the results of different
two-channel contributions. 
In this case  we need up
to 700 discretization points per channel adequately distributed between 0
and
70$a_\mu$. This relatively large number of points is necessary to
properly deal with the large final-state Coulomb interaction which could
be as large as $17e^2/\rho$ in the case of argon. 
Some $20$ to 30 discretization points per
channel would be enough for uncharged fragments in the initial and final
state. 

Next the calculations are repeated in the presence of polarization
potential.
It is more difficult to obtain convergence with the polarization
potential (\ref{eq:pol}) which is taken to be zero at small
distances below the cut off $\Lambda$. In this case to get numerical
convergence we have to integrate to very large distances $-$ up to
300$a_{\mu }$. We need up to 2000 discretization points per channel to
obtain
convergence.  More points are employed near the origin and
less at large distances.  For example, near the origin we take up to 60
equally spaced points per  unit length interval $a_\mu$; in the
intermediate region ($\rho = 15 - 30 a_\mu$) we take up to 8 equally
spaced points per unit length interval, and in the asymptotic
region ($\rho = 30 - 300 a_\mu$) we take up to 6 equally spaced points
per unit length interval.  

The short-range potential of the present problem extends to about ${\cal
R} = 20a_\mu$. It is customary to take the cut off $\Lambda$ of the
polarization potential much larger than the range ${\cal R}$ of the
short-range potential. For $\Lambda < {\cal R}$, the polarization
potential becomes much larger than the short-range potential and for
$\Lambda = 0$ it diverges.  The results for the cross section and the
transfer rates can be very sensitive to the value of the cut off for
$\Lambda <{\cal R}$. However, the polarization potential should be
effective in the
intermediate region: $\infty>\Lambda>{\cal R}$.  We find that for
$\infty>\Lambda>{\cal R}$, the results for the transfer rates are
very weakly dependent on $\Lambda$.  We considered the polarization
potential in this asymptotic region $\rho_1 > \Lambda \simeq 120a_\mu$. 
For a variation of $\Lambda$ in this region from $120a_\mu$ to about
160$a_\mu$, we find the transfer cross sections to be reasonably constant
and the reported transfer rates  of this study are  the averages of
these cross sections. If we increase $\Lambda$ beyond $160a_\mu$ the
effect of the polarization potential gradually decreases and transfer
rates gradually tend towards the rates calculated without the polarization
potential.

We present partial muon-transfer rates $\lambda^{\scriptsize
\mbox{tr}}_{nl}$ and total transfer rates $\lambda_{\scriptsize
\mbox{tot}}^{\scriptsize \mbox{tr}}$.  We calculate the low-energy
muon-transfer rates from ($p\mu$)$_{1s}$ and ($d\mu$)$_{1s}$ to
Ne$^{10+}$, S$^{16+}$, and Ar$^{18+}$.  We find in our calculation that in
all cases the transfer takes place predominantly to the angular momenta
states $l=0,1$ of the $n=6$ orbital of (Ne$_\mu)^{9+} $, $n=9$ orbital of
(S$_\mu)^{15+} $, and $n=10$ orbital of (Ar$_\mu)^{17+} $. The transition
to other states of (X$_\mu)^{(Z-1)+} $ is negligibly small.  
Hence in this
work we only present muon-transfer rates for the $l=0,1$ states of the
above orbitals of neon, sulphur, and argon. 
The low-energy partial rates $\lambda^{\scriptsize
\mbox{tr}}_{1s\rightarrow nl}$ $/10^{10}$ s$^{-1}$ and total rates
$\lambda_{\scriptsize \mbox{tot}}^{\scriptsize \mbox{tr}}$ $/10^{10}$
s$^{-1}$ with and without the polarization potential are presented in
Tables I to V together with the results of experimental works. In all
cases the partial transfer rates without the polarization potential
saturates to a
reasonably constant value for $E<0.01$ eV.  A similar behavior is also
observed in the presence of the polarization potential. In all cases
presented, the rate of transition to the s state of the muonic atomic
orbital is larger than that to the p state both in the presence and
absence of polarization potential. For example, in the case of Ne$^{10+}$,
the muon-transfer rate to the 6s state of (Ne$_\mu)^{9+}$ is larger than
that to the 6p state, and so on. 

First we consider 
the results in Table I for muon transfer from 
muonic protium
($p\mu$)$_{1s}$  to Ar$^{18+}$. In this case we find that a constant value
for the transfer rate is achieved for $E\le 0.04$ eV and we present
results
up to this energy.  The present total transfer rate of
$(12.9 \pm 0.4)\times 10^{10}$ s$^{-1}$ is in 
good agreement with experiments of Refs. \cite{pr97,jetp66,nci69}.
However, it disagrees strongly with experiments of Refs. 
\cite{nci67,np80}.
In the case of muon transfer from 
($d\mu$)$_{1s}$
to Ar$^{18+}$, we find in Table II that  the total transfer rate
has decreased in this case compared to that  in Table I, in agreement
with the observed experimental trend \cite{pr97}.
For the rates reported in Table II a constant value is obtained for
$E\le 0.04$ eV. 
The present total transfer rate of $(5.3\pm 0.4)\times 10^{10}$
s$^{-1}$ for $(d\mu)_{1s}$ to argon is in reasonable agreement with 
the experimental rates of Refs. \cite{pr97,nci67a}.

In the cases of muon-transfer rates from hydrogen isotopes to muonic
sulphur  and neon reported below in Tables III, IV, and V, a constant
value for the rate was not obtained till 
$E=0.04$ eV, as in the case of argon above. Hence in these cases
we report the results of our calculation for energies 0.01 and 0.04 eV
with the experimental energy lying in between. 
In Table III
the results  for muon transfer from 
muonic protium
($p\mu$)$_{1s}$  to S$^{16+}$ are shown.  
The present total transfer rate of
$(12.0 \pm 0.4)\times 10^{10}$ s$^{-1}$ is in 
reasonable agreement with the experimental rate $8.9\times
10^{10}$ s$^{-1}$ of Ref. \cite{jpb93}.
In the case of muon transfer from muonic deuterium
($d\mu$)$_{1s}$
to S$^{16+}$, we find from Table IV  that the total transfer rate
of $(12.7 \pm 0.4)\times 10^{10}$ s$^{-1}$ is also in good agreement
with the experimental rate $11.0\times 10^{10}$ s$^{-1}$
of Ref.  \cite{jpb93}. The transfer rate 
has increased in this case compared to that  in Table III for muon
transfer from protium   in agreement
with experimental observation \cite{jpb93}. Finally,
In Table V we exhibit  muon transfer from muonic
deuterium ($d\mu$)$_{1s}$ to Ne$^{10+}$.  The present total transfer
rate of $(8.4 \pm 0.2)\times 10^{10}$ s$^{-1}$ at 0.01 eV is in good
agreement with experimental rate $10.1 \times 10^{10}$ s$^{-1}$ 
of Ref. \cite{pra95}. 

From Tables I to V we find that in most cases the transition rates
calculated in the present model without
the polarization potential is already in reasonable agreement with
experiment. This suggests that the present model based on the
close-coupling approximation to coupled 
Faddeev-Hahn-equations is very suitable for muon transfer reactions
with
neon, sulphur, and argon. A similar conclusion can be made from
our previous study on  muon transfer reactions with hydrogen \cite{pra2000},
helium, lithium \cite{jphysb}, carbon, and oxygen \cite{pra2001}.
Although, the effect of polarization in these cases is expected to be 
large on the observables of the elastic-channel observables at low
energies, this effect is not found to be so pronounced on the
muon-transfer rates
calculated in this work.

\section{Conclusion}

We have calculated muon-transfer rates from muonic protium and deuterium
to bare nuclei neon, sulphur, and argon employing a full three-body
quantum-mechanical description of rearrangement scattering by solving the
Faddeev-Hahn-type equations \cite{hahn68} using close-coupling
approximation. The exact
inclusion of the final-state Coulomb interaction in the rearrangement
channel (without approximation) guarantees the correct asymptotic form of
the wave function in this channel.  We also included a
polarization potential at large distances in the initial channel beyond a
cut off distance $\Lambda$.  In these problems the short-range part of the
interaction extends to about 20 muonic atomic units ($25a_\mu$).  By
varying $\Lambda$ in the intermediate range between 120$a_\mu$ to
160$a_\mu$ we find that the transfer rates are almost independent of the
cut off $\Lambda$. This makes the reported transfer rates of this work
quasi-independent of cut off.  It is shown that in the present approach,
the application of a close-coupling-type ansatz leads to satisfactory
results for direct muon-transfer reactions from muonic hydrogen to bare
nuclei neon, sulphur, and argon.  The present rates are in good agreement
with experiments \cite{pr97,jetp66,nci69,nci67a,pra95,jpb93}.
Our
calculation also reproduces the observed experimental trend of transfer
rates to sulphur and argon from protium and deuterium \cite{pr97}. The
transfer rate increases with the decrease of mass of the hydrogen isotope
in the case of argon; the reverse is true for sulphur. 

It is interesting to note from the works of Refs. 
\cite{pra2000,jphysb,pra2001} and the present investigation that a
low-order approximation to the Faddeev-Hahn-type equations as considered
here produces very good muon-transfer rates from hydrogen isotopes to bare
nuclei at low energies. In previous studies it was noted that low-order
approximation to these equations also produces very accurate scattering
observables at low energies for electron-hydrogen \cite{Z} and
positron-hydrogen systems \cite{fb99a} in agreement with the variational
results. It is well-known that similar low-order approximation to the
Schr\"odinger equation leads usually to poor results at low energies
\cite{1,5,rmp62}. Hence low-order approximation to the Faddeev-Hahn-type
equations is a very attractive alternative for studying low-energy Coulomb
charge-transfer reactions in general.

\acknowledgments
We acknowledge the support from FAPESP (Funda\c{c}\~{a}o
de Amparo \~{a} Pesquisa do Estado de S\~{a}o Paulo) of  Brazil.
The numerical calculations have been performed on the IBM SP2
Supercomputer of the Departamento de
F\'\i sica - IBILCE - UNESP,
S\~{a}o Jos\'e do Rio Preto, Brazil.

\mediumtext

\begin{table}
{Table I. Low energy partial
$\lambda^{\scriptsize \mbox{tr}}_{1s\rightarrow
nl}$$/10^{10}$$\mbox{s}^{-1}$
and total $\lambda_{\scriptsize
\mbox{tot}}^{\scriptsize \mbox{tr}}/10^{10}$$\mbox{s}^{-1}$
muon transfer rates reduced to liquid-hydrogen density
$N_0 = 4.25\times10^{22}$ $\mbox{cm}^{-3}$
from muonic protium (${p\mu}$)$_{1s}$
to hydrogen-like excited state of muonic argon
(${{\mbox {Ar}}_\mu})^{17+}_{n=10}$ together with some experimental
results $\lambda_{\scriptsize
\mbox{ }}^{\scriptsize \mbox{tr}}$ $/10^{10}$ $\mbox{s}^{-1}$.}
\begin{tabular}{lcccccccclccccccc}
\multicolumn{1}{l}{Energy}                &
\multicolumn{1}{c}{ }                     &
\multicolumn{2}{c}{$U_{\scriptsize \mbox{pol}}(\rho_1) = 0$} &
\multicolumn{2}{c}{With polarization} &
\multicolumn{3}{c}{Experiment}\\
\multicolumn{1}{l}{$E$ (eV)}              &
\multicolumn{1}{l}{$(nl)$}                &
\multicolumn{1}{c}{$\lambda^{\scriptsize \mbox{tr}}_{1s\rightarrow nl}$} &
\multicolumn{1}{c}{$\lambda_{\scriptsize
\mbox{tot}}^{\scriptsize \mbox{tr}}$} &
\multicolumn{1}{c}{$\lambda^{\scriptsize \mbox{tr}}_{1s\rightarrow nl}$} &
\multicolumn{1}{c}{$\lambda_{\scriptsize
\mbox{tot}}^{\scriptsize \mbox{tr}}$} &
\multicolumn{3}{c}{$\lambda_
{\scriptsize ^1{\mbox H}_\mu+{\mbox Ar}}^{\scriptsize \mbox{tr}}$}  &
\\ \hline
$ 0.04 $ & $ 10s $ & $ 5.0\pm 0.2 $ & $ $ & $ 8.1 \pm 0.2 $ & 
& $ 16.3 \cite{pr97} $ &12.0\cite{jetp66} & 14.6 \cite{nci69}\\
$ $ & $ 10p $ & $ 3.9\pm 0.2 $ & $ 8.9\pm 0.4 $ & $ 4.8 \pm 0.2 $
& $ 12.9 \pm 0.4 $ 
& & \ \ 35\cite{nci67} & 98 \cite{np80}\\
\hline
$ 0.06 $
& $ 10s $ & $ 5.0 \pm 0.2 $ & $ $ & $ 7.9 \pm 0.2$
& $ $ & $ $ & $ $\\
$ $
& $ 10p $ & $ 3.9 \pm 0.1 $ & $ 8.9 \pm 0.3 $ & $ 4.6 \pm 0.1$
& $ 12.5 \pm 0.3 $\\
\hline
$ 0.1 $
& $ 10s $ & $ 4.8 \pm 0.1 $ & $ $ & $ 6.8 \pm 0.2$
& $ $ & $ $ & $ $\\
$ $
& $ 10p $ & $ 3.8 \pm 0.1 $ & $ 8.6 \pm 0.3 $ & $ 4.1 \pm 0.1$
& $ 10.9 \pm 0.3$\\
\end{tabular}
\end{table}

\mediumtext

\begin{table}

{Table II. Same as in Table I from muonic deuterium $(d\mu)_{1s}$  to 
muonic argon
(Ar$_\mu$)$^{17+}_{n=10}$}
\begin{tabular}{lcccccccclccccccc}
\multicolumn{1}{l}{Energy}                &
\multicolumn{1}{c}{ }                     &
\multicolumn{2}{c}{$U_{\scriptsize \mbox{pol}}(\rho_1) = 0$} &
\multicolumn{2}{c}{With polarization} &
\multicolumn{2}{c}{Experiment}\\
\multicolumn{1}{l}{$E$ (eV)}              &
\multicolumn{1}{l}{$(nl)$}                &
\multicolumn{1}{c}{$\lambda^{\scriptsize \mbox{tr}}_{1s\rightarrow nl}$} &
\multicolumn{1}{c}{$\lambda_{\scriptsize
\mbox{tot}}^{\scriptsize \mbox{tr}}$} &
\multicolumn{1}{c}{$\lambda^{\scriptsize \mbox{tr}}_{1s\rightarrow nl}$} &
\multicolumn{1}{c}{$\lambda_{\scriptsize
\mbox{tot}}^{\scriptsize \mbox{tr}}$} &
\multicolumn{3}{c}{$\lambda_
{\scriptsize ^2{\mbox H}_\mu+{\mbox Ar}}^{\scriptsize \mbox{tr}}$}  &
\\ \hline
$ 0.04 $
& $ 10s $ & $ 1.3 \pm 0.1 $ & $ $ & $ 3.4 \pm 0.2 $
& $  $ & $ $ & $ $\\
$ $
& $ 10p $ & $ 0.9 \pm 0.1 $ & $ 2.2 \pm 0.2 $ & $ 1.9 \pm 0.2 $
& $ 5.3 \pm 0.4 $ & $ 8.6 \cite{pr97}$ & $ 9.4 \cite{nci67a}$ \\
\hline
$ 0.06 $
& $ 10s $ & $ 1.2 \pm 0.1 $ & $ $ & $ 3.2 \pm 0.2$
& $ $ & $ $ & $ $\\
$ $
& $ 10p $ & $ 0.9 \pm 0.1 $ & $ 2.1 \pm 0.2 $ & $ 1.8 \pm 0.2 $
& $ 5.0 \pm 0.4 $\\
\hline
$ 0.1 $
& $ 10s $ & $ 1.1 \pm 0.1 $ & $ $ & $ 2.8 \pm 0.1$
& $ $ & $ $ & $ $\\
$ $
& $ 10p $ & $ 0.8 \pm 0.1 $ & $ 1.9 \pm 0.2 $ & $ 1.4 \pm 0.1 $
& $ 3.2 \pm 0.2 $\\
\end{tabular}
\end{table}
\mediumtext

\begin{table}
{Table III. Same as in Table I from muonic protium $(p\mu)_{1s}$  to 
muonic sulphur
(S$_\mu$)$^{15+}_{n=9}$}
\begin{tabular}{lcccccccccccccccc}
\multicolumn{1}{l}{Energy}                &
\multicolumn{1}{c}{ }                     &
\multicolumn{2}{c}{$U_{\scriptsize \mbox{pol}}(\rho_1) = 0$} &
\multicolumn{2}{c}{With polarization} &
\multicolumn{1}{c}{Experiment}\\
\multicolumn{1}{l}{$E$ (eV)}              &
\multicolumn{1}{l}{$(nl)$}                &
\multicolumn{1}{c}{$\lambda^{\scriptsize \mbox{tr}}_{1s\rightarrow nl}$} &
\multicolumn{1}{c}{$\lambda_{\scriptsize
\mbox{tot}}^{\scriptsize \mbox{tr}}$} &
\multicolumn{1}{c}{$\lambda^{\scriptsize \mbox{tr}}_{1s\rightarrow nl}$} &
\multicolumn{1}{c}{$\lambda_{\scriptsize
\mbox{tot}}^{\scriptsize \mbox{tr}}$} &
\multicolumn{1}{c}{$\lambda_
{\scriptsize ^1{\mbox H}_\mu+{\mbox S}}^{\scriptsize \mbox{tr}}$}  &
\\ \hline
$ 0.01 $
& $ 9s $ & $ 6.5 $ & $ $ & $ 8.2 \pm 0.2$
& $ $ & $ $ & $ $\\
$ $
& $ 9p $ & $ 3.1 $ & $ 9.6 $ & $ 3.8 \pm 0.2$
& $ 12.0 \pm 0.4 $ & $ 8.9 \cite{jpb93} $\\
\hline
$ 0.04 $
& $ 9s $ & $ 6.9 $ & $ $ & $ 8.4 \pm 0.2$
& $ $ & $ $ & $ $\\
$ $
& $ 9p $ & $ 3.4 $ & $ 10.3 $ & $ 4.1 \pm 0.2$
& $ 12.5 \pm 0.4$\\
\end{tabular}
\end{table}

\mediumtext

\begin{table}
{Table IV. Same as in Table I from muonic deuterium $(d\mu)_{1s}$  to 
muonic sulphur
(S$_\mu$)$^{15+}_{n=9}$}
\begin{tabular}{lcccccccccccccccc}
\multicolumn{1}{l}{Energy}                &
\multicolumn{1}{c}{ }                     &
\multicolumn{2}{c}{$U_{\scriptsize \mbox{pol}}(\rho_1) = 0$} &
\multicolumn{2}{c}{With polarization} &
\multicolumn{1}{c}{Experiment}\\
\multicolumn{1}{l}{$E$ (eV)}              &
\multicolumn{1}{l}{$(nl)$}                &
\multicolumn{1}{c}{$\lambda^{\scriptsize \mbox{tr}}_{1s\rightarrow nl}$} &
\multicolumn{1}{c}{$\lambda_{\scriptsize
\mbox{tot}}^{\scriptsize \mbox{tr}}$} &
\multicolumn{1}{c}{$\lambda^{\scriptsize \mbox{tr}}_{1s\rightarrow nl}$} &
\multicolumn{1}{c}{$\lambda_{\scriptsize
\mbox{tot}}^{\scriptsize \mbox{tr}}$} &
\multicolumn{1}{c}{$\lambda_
{\scriptsize ^2{\mbox H}_\mu+{\mbox S}}^{\scriptsize \mbox{tr}}$}  &
\\ \hline
$ 0.01 $
& $ 9s $ & $ 6.8 $ & $ $ & $ 7.9 \pm 0.2$
& $ $ & $ $ & $ $\\
$ $
& $ 9p $ & $ 4.0 $ & $ 10.8 $ & $ 4.8 \pm 0.2 $
& $ 12.7 \pm 0.4 $ & $ 11.0 \cite{jpb93} $\\
\hline
$ 0.04 $
& $ 9s $ & $ 8.7 $ & $ $ & $ 9.7 \pm 0.2$
& $ $ & $ $ & $ $\\
$ $
& $ 9p $ & $ 4.4 $ & $ 13.1 $ & $ 4.9 \pm 0.2 $
& $ 14.6 \pm 0.4 $\\
\end{tabular}
\end{table}
\mediumtext

\mediumtext

\begin{table}
{Table V. Same as in Table I from muonic deuterium $(d\mu)_{1s}$  to 
muonic neon
(Ne$_\mu$)$^{9+}_{n=6}$}
\begin{tabular}{lcccccccccccccccc}
\multicolumn{1}{l}{Energy}                &
\multicolumn{1}{c}{ }                     &
\multicolumn{2}{c}{$U_{\scriptsize \mbox{pol}}(\rho_1) = 0$} &
\multicolumn{2}{c}{With polarization} &
\multicolumn{1}{c}{Experiment}\\
\multicolumn{1}{l}{$E$ (eV)}              &
\multicolumn{1}{l}{$(nl)$}                &
\multicolumn{1}{c}{$\lambda^{\scriptsize \mbox{tr}}_{1s\rightarrow nl}$} &
\multicolumn{1}{c}{$\lambda_{\scriptsize
\mbox{tot}}^{\scriptsize \mbox{tr}}$} &
\multicolumn{1}{c}{$\lambda^{\scriptsize \mbox{tr}}_{1s\rightarrow nl}$} &
\multicolumn{1}{c}{$\lambda_{\scriptsize
\mbox{tot}}^{\scriptsize \mbox{tr}}$} &
\multicolumn{1}{c}{$\lambda_
{\scriptsize ^2{\mbox H}_\mu+{\mbox S}}^{\scriptsize \mbox{tr}}$}  &
\\ \hline
$ 0.01 $
& $ 6s $ & $ 6.9 $ & $ $ & $ 8.2 \pm 0.2$
& $ $ & $ $ & $ $\\
$ $
& $ 6p $ & $ <0.1 $ & $ 7.0 $ & $ <0.2 $
& $ 8.4 \pm 0.2 $ & $ 10.1 \cite{pra95} $\\
\hline
$ 0.04 $
& $ 6s $ & $ 4.7 $ & $ $ & $ 6.2 \pm 0.2$
& $ $ & $ $ & $ $\\
$ $
& $ 6p $ & $ <0.05 $ & $ 4.75 $ & $ <0.1 $
& $ 6.3 \pm 0.2 $\\
\end{tabular}
\end{table}

\end{document}